\documentclass[a4paper,12pt,oneside]{article}
\usepackage[latin1]{inputenc}
\usepackage{amsmath}
\usepackage{amsfonts}
\usepackage{amssymb}
\usepackage{graphicx}
\usepackage{footnote}
\usepackage{float}
\usepackage{subfigure}
\DeclareGraphicsExtensions{.bmp,.png,.pdf,.jpg,.eps}

\usepackage[colorlinks]{hyperref}
\hypersetup{
    colorlinks=true,
    linkcolor=blue,
    filecolor=magenta,  
    urlcolor=[rgb]{0.00,0.00,0.50},    
    citecolor=red,
    }

\usepackage{url}

\advance \textheight by \topskip
\usepackage{amsmath}
\usepackage[english]{babel}
\makeatletter
 \@addtoreset{equation}{section}
\makeatother

\usepackage{amsmath}
\numberwithin{equation}{section}

\advance \textheight by \topskip
\usepackage{amsmath}
\usepackage[english]{babel}

\def\be{\begin{equation}}
\def\ee{\end{equation}}

\def\A{\mathbb A}

\def\Z{\mathbb Z}
\def\R{\mathbb R}

\def\I{\mathbb{I}}

\usepackage{todonotes}

\begin{document}

\title{
{\bf 
Hidden superconformal symmetry:\\ Where does it come from?
 }}

\author{{\bf  Luis Inzunza and Mikhail S. Plyushchay} 
 \\
[8pt]
{\small \textit{ Departamento de F\'{\i}sica,
Universidad de Santiago de Chile, Casilla 307, Santiago 2,
Chile  }}\\
[4pt]
 \sl{\small{E-mails:   
 \textcolor{blue}{luis.inzunza@usach.cl},
\textcolor{blue}{mikhail.plyushchay@usach.cl}
}}
}
\date{}
\maketitle

\begin{abstract}
It is  known that a single quantum harmonic oscillator
is characterized by a hidden spectrum generating superconformal symmetry,
but its origin has remained rather obscure.
We show how this hidden superconformal  symmetry can be derived 
by applying a nonlocal
Foldy-Wouthuysen transformation   to 
three   extended  systems with 
fermion degrees of freedom.
The associated systems have essentially different 
nature from the point of view of conventional  supersymmetric 
quantum mechanics, and generate the desired hidden symmetry 
in three different ways. 
We also trace out 
how the  hidden 
superconformal symmetry of  the 
quantum free particle system is produced
in the limit of zero frequency.
\end{abstract}

\vskip.5cm\noindent

\section{Introduction}\label{Sec1}

Supersymmetry relates bosons and fermions on the
basis of  $\Z_2$-graded  superalgebras.
Supersymmetry 
in quantum mechanics  is implemented by
separating   symmetry generators 
into ``even" (``bosonic") and ``odd" (``fermionic") subsets.
Coherently with this, the Hamiltonian's eigenstates 
are separated  into ``bosonic" and ``fermionic" states.
A standard implementation of supersymmetry in 
quantum mechanics is realized by introducing 
a set of fermion degrees of freedom in addition to
the bosonic degrees of freedom described by usual coordinate and momentum 
operators.
The fermionic set is realized by matrices of finite dimension,
or in terms of Grassmann variables in 
superspace formulation of supersymmetric quantum mechanics
\cite{Witten,CoKhSu}.

There exist, however, quantum mechanical systems  
in which supersymmetry is realized without 
introducing  additional degrees of freedom.
 In such non-extended systems,  space reflection
can  play a role of the 
$\Z_2$-grading operator \cite{Ply1,CorPly,CJNP}. 
The most known example of such a system corresponds, probably,  to 
a one-dimensional quantum harmonic oscillator,
in which there appears a  spectrum generating hidden superconformal 
$\mathfrak{osp}(1|2)$ symmetry \cite{CroRit,BalSchBar,CarPly,BCLW}. 
This hidden superconformal symmetry
has a further peculiarity in comparison with
Witten's supersymmetric quantum mechanics\,:
all its odd generators
have a dynamical character 
being explicitly depending on time 
integrals of motion of the system. 

Thus, a hidden superconformal symmetry of the quantum harmonic oscillator
has a rather obscure origin and  nature  from a perspective
of conventional realization of supersymmetric quantum mechanics. 

In the present paper we 
derive 
the hidden superconformal $\mathfrak{osp}(1|2)$ symmetry
of the quantum harmonic oscillator and its nonlocal $\mathfrak{osp}(2|2)$ extension
by applying a peculiar nonlocal 
Foldy-Wouthuysen transformation
to three different associated systems 
with conventional
quantum mechanical matrix fermion degrees of freedom.
We also investigate   
a  rather nontrivial transformation
of the hidden superconformal 
symmetry of the harmonic oscillator 
into the corresponding superconformal symmetry of  the 
quantum free particle system.

The paper is organized as follows.
In Section \ref{Sec2} we describe the 
hidden superconformal $\mathfrak{osp}(1|2)$ symmetry
of the quantum harmonic oscillator and its nonlocal $\mathfrak{osp}(2|2)$ extension. 
In Section \ref{Sec3} we consider an extended system with 
fermionic degrees of freedom that represents a doubled 
quantum harmonic oscillator. Such a system  cannot be generated
within a framework of a usual construction of supersymmetric quantum mechanics 
with supercharges anti-commuting for the Hamiltonian.
The extended system  possesses the super-extended Schr\"odinger 
symmetry  which under appropriate nonlocal 
Foldy-Wouthuysen transformation reduces and reproduces
the hidden superconformal symmetry of the quantum harmonic 
oscillator. 
In Section \ref{Sec4} we show how the hidden superconformal
symmetry of the quantum oscillator can also be produced via 
the dual Darboux transformations.
In Section \ref{Sec5} we consider 
the extended system obtained via a  one-parametric 
two-step  Darboux transformation of the quantum harmonic
oscillator. Such a  quantum system corresponds 
to the anomaly-free scheme of quantization of a classical system
with the second-order supersymmetry  and is described by 
a nonlinear super-extended Schr\"odinger symmetry. 
In the appropriate limit it  reproduces the system from Section  \ref{Sec3}
in which time-independent supercharges anti-commute 
for the central charge of its Lie type super-extended Schr\"odinger 
symmetry.  In Section \ref{Sec6} we 
consider a rather peculiar zero  frequency limit applied 
to the  system  from Section \ref{Sec3} and 
show how the hidden superconformal symmetry
of the free quantum particle is generated.
Section \ref{Sec7} is devoted to the discussion  and outlook.

\section{Hidden superconfromal symmetry of the quantum harmonic oscillator}\label{Sec2}
 
  In a system of units with Plank constant $\hbar=1$, 
  frequency  $\omega=1$ 
  and  mass  $m=\frac{1}{2}$, 
 Hamiltonian  of the quantum harmonic oscillator is
\begin{equation}\label{Lqhosc}
L=-\frac{d^2}{dx^2}+x^2,
\end{equation} 
and its  spectrum is given by a discrete set  $E_n=2n+1$, $n=0,1,\ldots$. 
 We take  ladder operators  in a form
\begin{equation}
a^{\pm}=\mp\frac{d}{dx}+x,\qquad [a^+,a^-]=2\,.
\end{equation}
They anticommute with reflection operator
$\mathcal{R}$ defined by  $\mathcal{R}^2=1$, 
$\mathcal{R}x=-x\mathcal{R}$, and their
anti-commutator generates  the Hamiltonian,
$\{a^+,a^-\}=2L$. 
Taking $\mathcal{R}$ as a $\Z_2$-grading operator, 
we identify the $a^\pm$ as odd, fermionic operators,
$\{\mathcal{R},a^\pm\}=0$.
Hamiltonian $L$ and quadratic  operators
$(a^{\pm})^2=\frac{d^2}{dx^2}+x^2+1 \mp 2 x\frac{d}{dx}$
are identified then as even, bosonic operators,
$[\mathcal{R},L]=0$, $[\mathcal{R},(a^\pm)^2]=0$.
This set of operators generates a Lie superalgebra
in which commutators of even and odd 
generators with Hamiltonian  are nontrivial\,:
$ [L,a^\pm]=\pm 2a^{\pm}$, 
$ [L,(a^\pm)^2]=\pm 4(a^{\pm})^2$.
Consequently, operators $a^\pm$ and $(a^\pm)^2$  are not integrals of
motion in the sense of Heisenberg equations of motion
$\frac{d}{dt}A=\frac{\partial}{\partial t}A-i[A,L]$.
The  operators   ``dressed"  by 
a unitary evolution operator
$U(t)=\exp (iLt)$ give the corresponding
even and odd explicitly depending on time integrals of motion 
$U^{-1}(t)a^\pm U(t)$
and $U^{-1}(t)(a^\pm)^2 U(t)$.
We shall refer to integrals of such a nature as to dynamical integrals.
Introducing 
 the rescaled operators
\begin{equation}\label{localGen}
J_0=\frac{1}{4}L,\qquad 
J_\pm=\frac{1}{4}e^{\mp4it}(a^{\pm})^2\,,\qquad \alpha_{\pm}=\frac{1}{4}e^{\mp i2t}a^{\pm}\,,
\end{equation}
we obtain the superalgebra with
nontrivial (anti)-commutation relations
\begin{equation}
\label{sl2}
[J_0,J_\pm]=\pm J_\pm,\qquad [J_-,J_+]=2J_0\,,
\end{equation} 
\begin{equation}
\label{osp(21.1)}
\{\alpha_{+},\alpha_-\}=\frac{1}{2}J_0\,,
\qquad\{\alpha_{\pm},\alpha_\pm\}=\frac{1}{2}J_\pm\,,
\end{equation}
\begin{equation}
\label{osp(21.3)}
[J_0,\alpha_{\pm}]=\pm\frac{1}{2}\alpha_{\pm},\qquad
[J_\pm,\alpha_{\mp}]=\mp\alpha_{\pm}\,.
\end{equation}
The superalgebra (\ref{sl2}), (\ref{osp(21.1)}),  (\ref{osp(21.3)}) 
describes the hidden superconformal 
$\mathfrak{osp}(1|2)$ symmetry of the quantum harmonic 
oscillator \cite{CroRit,BalSchBar}.
The set of even integrals  $J_0$, $J_\pm$ generates
the $\mathfrak{sl}(2,\R)$ subalgebra (\ref{sl2}), and  relations
(\ref{osp(21.3)}) 
mean that fermionic generators $\alpha_\pm$
form  a spin-$1/2$ representation  of this Lie sub-algebra.
All the generators of the $\mathfrak{osp}(1|2)$ superconformal 
algebra are local in $x$ operators. 
The grading operator $\mathcal{R}$, being  even time-independent integral 
of motion, can be presented in a form 
$\mathcal{R}=\sin (\frac{\pi}{2}L)$ which 
explicitly reveals its non-local nature.
Expanding the set of local integrals (\ref{localGen})
by nonlocal time-independent even integral $\mathcal{R}$ and 
by dynamical odd integrals 
\begin{equation}\label{beta}
\beta_{\pm}=i\mathcal{R}\alpha_{\pm}\,,
\end{equation}
we extend the $\mathfrak{osp}(1|2)$ 
for $\mathfrak{osp}(2|2)$ superconformal algebra in which 
we have additionally the following 
nontrivial (anti)-commutation relations which
involve the non-local integrals $\mathcal{R}$ and $\beta_\pm$\,:
\begin{equation}
\label{beta1}
[J_0,\beta_{\pm}]=\pm\frac{1}{2}\beta_{\pm}\,,\qquad
[J_{\pm},\beta_{\mp}]=\mp\beta_\pm\,,
\ee
\be
\qquad\{\beta_{\pm},\beta_{\pm}\}=\frac{1}{2}J_{\pm}\,,\qquad \{\beta_{+},\beta_{-}\}=\frac{1}{2}J_0\,,
\qquad
\{\alpha_{\pm},\beta_{\mp}\}=\mp\frac{i}{2}Z\,,
\ee
\be
[Z,\alpha_{\pm}]=\frac{i}{2}\beta_{\pm}\,,
\qquad [Z,\beta_{\pm}]=-\frac{i}{2}\alpha_\pm\,,
\end{equation}
where we introduced a notation
\begin{equation}
Z=-\frac{1}{4}\mathcal{R}\,.
\end{equation} 
In terms  of linear combinations
 \begin{equation}
\gamma_\pm=\alpha_\pm + i\beta_\pm\,,\qquad 
\delta_\pm=\gamma_\mp^\dagger=\alpha_\pm - i\beta_\pm\,,
\end{equation}
 a part of superalgebra involving 
odd generators can be  presented in an alternative form
\begin{equation}
[J_0,\gamma_\pm]=\pm\frac{1}{2}\gamma_\pm\,,
\qquad
[J_0,\delta_\pm]=\pm\frac{1}{2}\delta_\pm\,,
\qquad
[J_\pm,\gamma_\mp]=\mp\gamma_\pm\,,
\qquad
[J_\pm,\delta_\mp]=\mp\delta_\pm\,,
\end{equation}
\begin{equation}
[Z,\gamma_\pm]=\frac{1}{2}\gamma_\pm\,,
\qquad
[Z,\delta_\pm]=-\frac{1}{2}\delta_\pm\,,
\qquad
\{\gamma_\pm,\delta_\pm\}=J_\pm\,,
\qquad
\{\gamma_\pm,\delta_\mp\}=J_0\mp Z\,.
\end{equation}

\section{Extended system with super-Schr\"odinger symmetry and 
nonlocal Foldy-Wouthuysen transformation}\label{Sec3}

Now let us show how the described hidden superconformal 
$\mathfrak{osp}(1|2)$ and $\mathfrak{osp}(2|2)$ symmetries  
 of a single oscillator 
can be 
`extracted' from the super-Schr\"odinger 
symmetry of the
extended quantum harmonic oscillator system
described by the Hamiltonian
\be\label{Hextbr}
\mathcal{H}=
\left(
\begin{array}{cc}
L &  0 \\
 0&   L     
\end{array}
\right).
\ee
This system represents two copies of the
quantum harmonic oscillator (\ref{Lqhosc}),
and has three obvious matrix integrals of motion 
given by the Pauli matrices
$\sigma_1$, $\sigma_2$ and $\sigma_3$.
It is natural to identify the diagonal matrix 
$\Gamma=\sigma_3$ as a $\Z_2$-grading 
operator. Then Hamiltonian (\ref{Hextbr}) is identified as 
even generator and the anti-diagonal integrals
$\sigma_a$, $a=1,2$,  can be considered as odd supercharges. 
The peculiarity of the system (\ref{Hextbr}) is that
the supercharges $\sigma_a$  anticommute not 
for Hamiltonian but for central element, 
$\{\sigma_a,\sigma_b\}=2\delta_{ab}\mathbb{I}$,
$\mathbb{I}=\text{diag}\,(1,1)$.
All the energy levels of the extended 
system (\ref{Lqhosc}) including the lowest nonzero energy level 
 $E_0=1>0$ are  doubly degenerate, and  the Witten index 
 of this extended system
 equals zero. 
The doublet  of states corresponding to the lowest
 energy level is not annihilated by
 supercharges $\sigma_a$, 
 and (\ref{Lqhosc}) 
 is identified as a quantum 
 mechanical supersymmetric system in a phase 
 of spontaneously broken supersymmetry.
 The  time-independent even,
 $\mathcal{H}$ and $\sigma_3$,  
 and odd, $\sigma_1$ and $\sigma_2$,  integrals 
 of motion together with even central charge $\mathbb{I}$
 are local operators. Besides them,  the system (\ref{Hextbr})
 also has local dynamical integrals of motion
 \begin{equation}\label{J+-t}
\qquad \mathcal{J}_\pm=\frac{1}{4}e^{\mp i4t}\left(
\begin{array}{cc}
(a^{\pm})^2 &  0 \\
 0&     (a^{\pm})^2   
\end{array}
\right)=\left(
\begin{array}{cc}
J_\pm &  0 \\
 0&   J_\pm      
\end{array}
\right),
\end{equation}
\begin{equation}\label{C+-t}
\mathcal{C}_\pm=\frac{1}{4}e^{\mp i2t}\left(
\begin{array}{cc}
  a^{\pm}& 0  \\
 0 &   a^{\pm}
\end{array}
\right)=\left(
\begin{array}{cc}
  \alpha_\pm& 0  \\
 0 &   \alpha_\pm
\end{array}
\right),
\end{equation}
\begin{equation}
\mathcal{Q}_\pm=\frac{1}{4}e^{\mp i2t}
\left(
\begin{array}{cc}
  0&   a^\pm  \\
 a^\pm&   0     
\end{array}
\right)=
\left(
\begin{array}{cc}
  0&   \alpha_\pm  \\
 \alpha_\pm &   0     
\end{array}
\right),
\qquad \mathcal{S}_\pm=i\sigma_3\mathcal{Q}_\pm.
\end{equation}
Diagonal operators $\mathcal{J}_\pm$ and 
$\mathcal{C}_\pm$ are identified 
here as even generators, and 
antidiagonal dynamical integrals $\mathcal{Q}_\pm$
and $\mathcal{S}_\pm$ are odd generators.  
All these integrals generate the superalgebra
with the following  
 (anti)-commutation relations\,:
\begin{equation}
\label{supmat1}
[\mathcal{J}_0,\mathcal{J}_\pm]=
\pm \mathcal{J}_\pm\,,
\qquad [\mathcal{J}_-,
\mathcal{J}_+]=2\mathcal{J}_0\,,
\end{equation} 
\begin{equation}\label{supmat2}
[\mathcal{J}_0,\mathcal{C}_\pm]=\pm\frac{1}{2}\mathcal{C}_\pm\,,
\qquad
[\mathcal{J}_\pm,\mathcal{C}_\mp]=\mp\mathcal{C}_\pm\,,
\qquad
[\mathcal{C}_-,\mathcal{C}_+]=\frac{1}{2}\mathcal{I}\,,
\ee
\be\label{supmat2+}
[\mathcal{J}_0,\mathcal{Q}_\pm]=\pm\frac{1}{2}\mathcal{Q}_\pm\,,\quad
[\mathcal{J}_0,\mathcal{S}_\pm]=\pm\frac{1}{2}\mathcal{S}_\pm\,,\quad
[\mathcal{J}_\pm,\mathcal{Q}_\mp]=\mp \mathcal{Q}_\pm\,,\quad
[\mathcal{J}_\pm,\mathcal{S}_\mp]=\mp \mathcal{S}_\pm\,,
\ee
\begin{equation}\label{supmat3}
\{\Sigma_a,\Sigma_b\}=2\delta_{ab}\,\mathcal{I}\,,
\qquad 
\{\Sigma_1,\mathcal{Q}_\pm\}=\mathcal{C}_\pm\,,
\qquad
\{\Sigma_2,\mathcal{S}_\pm\}=\mathcal{C}_\pm\,,
\end{equation}
\be\label{supmat4}
\{\mathcal{Q}_\pm,\mathcal{Q}_\pm\}=\frac{1}{2}\mathcal{J}_\pm\,,
\quad
\{\mathcal{Q}_+,\mathcal{Q}_-\}=\frac{1}{2}\mathcal{J}_0\,,\quad
\{\mathcal{S}_\pm,\mathcal{S}_\pm\}=\frac{1}{2}\mathcal{J}_\pm\,,
\quad
\{\mathcal{S}_+,\mathcal{S}_-\}=\frac{1}{2}\mathcal{J}_0\,,
\ee
\be\label{supmat5}
\{\mathcal{Q}_+,\mathcal{S}_-\}=-\frac{i}{2}\mathcal{Z}\,,
\qquad
\{\mathcal{Q}_-,\mathcal{S}_+\}=\frac{i}{2}\mathcal{Z}\,,
\ee
\begin{equation}\label{supmat6+}
[\mathcal{Z},\Sigma_a]=\frac{i}{2}\epsilon_{ab}\Sigma_b\,,\qquad
[\mathcal{Z},\mathcal{Q}_{\pm}]=\frac{i}{2}\mathcal{S}_{\pm}\,,
\qquad [\mathcal{Z},\mathcal{S}_{\pm}]=-\frac{i}{2}\mathcal{Q}_\pm\,,
\ee
\be\label{supmat6}
[\mathcal{C}_\pm,\mathcal{Q}_\mp]=\mp\frac{1}{4} \Sigma_1,\qquad
[\mathcal{C}_\pm,\mathcal{S}_\mp]=\mp\frac{1}{4} \Sigma_2\,,
\end{equation}
where $\mathcal{J}_0$ is a rescaled Hamiltonian (\ref{Hextbr}), 
\begin{equation}\label{mathcalJ0}
\mathcal{J}_0=\frac{1}{4}\mathcal{H}=
\left(
\begin{array}{cc}
J_0 &  0 \\
 0&   J_0      
\end{array}
\right),
\end{equation}
and we introduced the notation
\begin{equation}\label{Sigmadef}
\Sigma_1=\frac{1}{2}\sigma_1\,,\qquad 
\Sigma_2=-\frac{1}{2}\sigma_2\,, \qquad
\mathcal{Z}=-\frac{1}{4}\sigma_3\,,
\qquad 
\mathcal{I}=\frac{1}{4}\mathbb{I}\,.
\end{equation} 
The not shown  (anti)commutators between generators are equal to zero.
By comparing this superalgebra and the structure of its 
generators with superalgebra and generators 
of the hidden superconformal $\mathfrak{osp}(1|2)$ 
and $\mathfrak{osp}(2|2)$ 
symmetries of the quantum harmonic oscillator,
it is obvious that the matrix 
integrals $\mathcal{J}_0$, $\mathcal{J}_\pm$, $\mathcal{Z}$,
$\mathcal{Q}_\pm$, $\mathcal{S}_\pm$
of the extended system (\ref{Hextbr})
are analogous to the corresponding 
integrals $J_0$, $J_\pm$, $Z$,
$\alpha_\pm$, $\beta_\pm$
of the quantum harmonic oscillator.
Because of the extension, the nonlocal integrals 
$Z$ and $\beta_\pm$ of the system 
(\ref{Lqhosc}) are changed here for
the corresponding local matrix integrals 
$\mathcal{Z}$ and $\mathcal{S}_\pm$.
The anti-commutator of additional fermionic integrals $\Sigma_a$ 
 with $\Sigma_b$ generates
a central charge $\mathcal{I}$, and 
 via the anti-commutators 
with odd dynamical integrals $\mathcal{Q}_\pm$ and
$\mathcal{S}_\pm$ they produce additional 
bosonic integrals $\mathcal{C}_\pm$,
see Eq. (\ref{supmat3}). 
The superalgebra (\ref{supmat1})--(\ref{supmat6})
 represents a super-extended Schr\"odinger 
symmetry of the matrix system (\ref{Hextbr})
with relations (\ref{supmat1}), (\ref{supmat2+}),
(\ref{supmat4}) and (\ref{supmat5}) corresponding 
to $\mathfrak{osp}(2|2)$ sub-superalgebra.

{}The comparison of the symmetries and generators 
of the systems  (\ref{Hextbr}) and (\ref{Lqhosc}) 
indicates  that the local $\mathfrak{osp}(1|2)$ 
and nonlocal $\mathfrak{osp}(2|2)$ hidden superconformal
symmetries of the quantum harmonic oscillator 
can be obtained by a certain projection (reduction)
of the local symmetries of the matrix system (\ref{Hextbr}).
To find the exact relation between these two systems and 
their symmetries, 
we apply to the extended system a unitary transformation 
$\mathcal{O}\mapsto \widetilde{\mathcal{O}}=U\mathcal{O}U^\dagger$
generated by the nonlocal matrix operator
\begin{equation}\label{Utrans}
U=U^\dagger=U^{-1}=\frac{1}{2}
\left(
\begin{array}{cc}
 1+\mathcal{R} &  1-\mathcal{R}  \\
  1-\mathcal{R}&   1+\mathcal{R}      
\end{array}
\right).
\end{equation}
This transformation acts in the following way on the basic
operators of the matrix system  (\ref{Hextbr})\,: 
$\widetilde{x}=x\sigma_1$, $\widetilde{p}=p\sigma_1$,
  $\widetilde{\sigma_1}=\sigma_1$, $\widetilde{\sigma_2}=\sigma_2\mathcal{R}$ 
  and $\widetilde{\sigma_3}=\sigma_3\mathcal{R}$, and 
 we also have  $\widetilde{\mathcal{R}}=\mathcal{R}$.
 As a consequence,  the central element $\mathcal{I}$ and 
 generators of the   $\mathfrak{sl}(2,\R)$ subalgebra, $\mathcal{J}_0$ and 
 $\mathcal{J}_\pm$,  do not change under this transformation, 
 while other generators take the following form\,:
\begin{equation}
\widetilde{Z}=\frac{1}{4}\left(
\begin{array}{cc}
  -\mathcal{R}& 0    \\
 0&   \mathcal{R}     
\end{array}
\right)\,,
\end{equation} 

\begin{equation}
\widetilde{\mathcal{Q}_\pm}=
\left(
\begin{array}{cc}
  \alpha_\pm& 0    \\
 0&   \alpha_\pm     
\end{array}
\right),\qquad
\widetilde{\mathcal{S}_\pm}=
\left(
\begin{array}{cc}
  i\mathcal{R}\alpha_\pm& 0    \\
 0&   -i\mathcal{R}\alpha_\pm     
\end{array}
\right)
=\left(
\begin{array}{cc}
  \beta_\pm& 0    \\
 0&   -\beta_\pm     
\end{array}
\right),
\end{equation}
\be
\widetilde{\Sigma_1}=\frac{1}{2}\sigma_1\,,\qquad
\widetilde{\Sigma_2}=-\frac{1}{2}\sigma_2\mathcal{R}\,,
\qquad
\widetilde{\mathcal{C}_\pm}=\sigma_1\alpha_\pm \,.
\ee
{}The unitary transformation generated by nonlocal operator
(\ref{Utrans})
diagonalizes the dynamical odd integrals 
$\mathcal{Q}_\pm$ and 
$\mathcal{S}_\pm$ which initially
have had the anti-diagonal form.
The transformation therefore is of a nature 
of Foldy-Wouthuysen transformation for
Dirac particle in external electric and magnetic fields \cite{FW}. 
It is interestingly to  note that the transformed odd and even integrals 
$\widetilde{\mathcal{Q}_\pm}$ and $\widetilde{\mathcal{C}_\pm}$
take the form of the original even and odd integrals 
$\mathcal{C}_\pm$  and $\mathcal{Q}_\pm$, respectively.
The transformed even, $\widetilde{\mathcal{Z}}$, and odd, 
$\widetilde{\mathcal{S}_\pm}$, generators of the superconformal 
$\mathfrak{osp}(2,2)$ sub-superalgebra of the 
super-extended Schr\"odinger symmetry of the system (\ref{Hextbr}) 
take a  nonlocal form.
We can reduce (or, in other words, project) the transformed system 
and its symmetries 
to the proper subspace of eigenvalue $+1$ of the matrix
$\sigma_3=\widetilde{\mathcal{R}\sigma_3}$ which corresponds,
according to Eq. (\ref{mathcalJ0}), 
to the single (non-extended) quantum harmonic oscillator system.
This can be done by multiplying all the transformed generators 
$\widetilde{\mathcal{X}_i}$
of the super-extended Schr\"odinger symmetry  from both sides
by the projector $\Pi_+=\frac{1}{2}(1+\sigma_3)$\,:
$\widetilde{\mathcal{X}_i}\mapsto  \Pi_+\widetilde{\mathcal{X}_i}\Pi_+$.
Since the transformed generators $\widetilde{\Sigma_{a}}$ and 
$\widetilde{\mathcal{C}_\pm}$ are anti-diagonal and anticommute 
with $\sigma_3$,
we loose them in the reduction procedure by mapping them 
into zero. Since  the central element $\mathcal{I}$  is 
generated in superalgebra of the  super-extended Schr\"odinger symmetry
of the system  (\ref{Hextbr}) via the anti-commutators of 
$\widetilde{\Sigma_{a}}$ with $\widetilde{\Sigma_{b}}$ and commutator of 
$\widetilde{\mathcal{C}_+}$ and $\widetilde{\mathcal{C}_-}$,
we also loose it as a generator of the surviving part
of the superalgebra.
The rest of the generators  
in the proper subspace of eigenvalue $+1$ of  
$\sigma_3$ will take exactly the form 
of the corresponding generators of the 
hidden superconformal $\mathfrak{osp}(2,2)$ symmetry 
of the quantum harmonic oscillator system 
(\ref{Lqhosc}).

\section{Superconformal symmetry via dual
Darboux transformations}\label{Sec4}

Though the extended system (\ref{Hextbr})  
has allowed us to derive
the hidden superconformal symmetry of the 
quantum harmonic oscillator, it cannot be generated directly 
within the framework of  supersymmetric quantum mechanics
with its underlying structure of Darboux transformations.
Nevertheless, we shall show in this section  how  the  super-extended 
Schr\"odinger symmetry can  be generated via a usual  supersymmetric 
extension of the quantum harmonic oscillator, and
 trace out  a difference in the associated  reduction procedure
leading to the hidden superconformal symmetry~\footnote{
See also refs. \cite{BecHus,BDH}  for the discussion of superconformal symmetry 
of the super-extended quantum harmonic oscillator  obtained
within the framework of conventional supersymmetric 
quantum mechanics construction.}.

The harmonic oscillator corresponds  to  the simplest case of the 
duality induced by Darboux transformations \cite{CIP}. This means 
that the same, modulo an additive shift,  superpartner 
can be produced  for a given system 
by choosing different states as  seed states
to generate a Darboux transformation. 
The presence of a nontrivial relative shift
will play, as we shall see,  a crucial role in generating
the structure of  super-extended Schr\"odinger symmetry.

Let us construct the Darboux intertwining operators
for the quantum harmonic oscillator by taking
its ground state    $\psi_0=Ce^{-x^2/2}$
as a seed state, where
a concrete value of a normalization constant $C$ is of no importance.
 They are nothing else as the ladder operators, 
\begin{equation}\label{A0Darboux}
A_0^-=\psi_0\frac{d}{dx}\frac{1}{\psi_0}=a^{-}, \qquad A_0^+=(A_0^-)^\dagger=a^+ \,,
\end{equation}
which  generate 
two mutually shifted copies of the quantum harmonic oscillator\,:
 $A_0^{+}A_0^-=a^+ a^-=L-1= H_-$ and 
$A_0^{-}A_0^+=a^{-}a^+=L+1= H_+$. 
The superpartner system for $H_-$ is therefore the same but shifted 
harmonic oscillator $H_+=H_-+2$. 
The operators $A_0^\pm=a^\pm$ intertwine the super-partner systems
$H_-$ and $H_+$,
\be\label{AAHH}
A^-_0H_-=H_+A^-_0\,, 
\qquad 
A^+_0H_+=H_-A^+_0\,.
\ee
Together $H_-$ and $H_+$ constitute the extended Hamiltonian
which can be presented in terms of a superpotential 
$W=-(\ln \psi_0)'=x$, 
\begin{equation}\label{hatH}
\widehat{\mathcal{H}}=-\frac{d^2}{dx^2}+W^2+\sigma_3W'=\left(
\begin{array}{cc}
  H_+& 0    \\
 0&  H_-     
\end{array}
\right)\,.
\end{equation}
The intertwining operators constitute the building blocks 
for  time-independent supercharges 
for the system (\ref{hatH}), 
\begin{equation}\label{Qasuper}
\widehat{\mathcal{Q}}_1=
\left(
\begin{array}{cc}
  0 & a^-    \\
 a^+&   0     
\end{array}
\right) ,\qquad
\widehat{\mathcal{Q}}_2=i\sigma_3\widehat{\mathcal{Q}}_1\,,
\end{equation}
$[\widehat{\mathcal{H}},\widehat{\mathcal{Q}}_a]=0$,
$\{\widehat{\mathcal{Q}}_a,\widehat{\mathcal{Q}}_b\}=2\delta_{ab}\widehat{\mathcal{H}}$,
with integral $\Gamma=\sigma_3$ identified as a $\Z_2$-grading operator.
Since a singlet ground state $\Psi_0=(0,\psi_0)^T$ 
of $\widehat{\mathcal{H}}$  is annihilated by both supercharges
$\widehat{\mathcal{Q}}_a$, the system  (\ref{hatH}), unlike 
(\ref{Hextbr}), corresponds to the case of exact, unbroken supersymmetry.

Instead of the ground state $\psi_0$, we can  take
a non-normalizable (non-physical) eigenstate 
$\psi_0^{-}=1/\psi_0=C^{-1}e^{x^2/2}$ of eigenvalue $E_{-0}=-1$ 
of the quantum harmonic oscillator
$L$ to generate the Darboux transformation.
The corresponding  operators in this case are
\begin{equation}\label{A-+0def}
A_{-0}^-=
\psi_0^-\frac{d}{dx}\frac{1}{\psi_{0}^-}=-a^{+},\qquad A_{-0}^+=
(A_{-0}^-)^\dagger=-a^{-}\,.
\end{equation}  
They satisfy relations $A_{-0}^+A_{-0}^{-}=L+1=H_+$, $A_{-0}^-A_{-0}^{+}=L-1=H_-$, 
$A_{-0}^-H_+=H_-A_{-0}^-$, $A_{-0}^+H_-=H_+A_{-0}^+$,
and generate a supersymmetric system described 
by the Hamiltonian operator
\begin{equation}\label{breveH}
\breve{\mathcal{H}}=-\frac{d^2}{dx^2}+W^2-\sigma_3W'=\left(
\begin{array}{cc}
  H_-& 0    \\
 0&  H_+     
\end{array}
\right)\,,
\end{equation}
which has two conserved supercharges
\be\label{Sasuper}
\breve{\mathcal{S}}_1=
\left(
\begin{array}{cc}
  0 & a^+    \\
 a^-&   0     
\end{array}
\right),\qquad
\breve{\mathcal{S}}_2=i\sigma_3\breve{\mathcal{S}}_1,
\end{equation}
$[\breve{\mathcal{H}},\breve{\mathcal{S}}_a]=0$, 
$\{\breve{\mathcal{S}}_a,\breve{\mathcal{S}}_b\}=2\delta_{ab}\breve{\mathcal{H}}$.
Hamiltonian (\ref{breveH}) and its supercharges (\ref{Sasuper}) are  
related to Hamiltonian (\ref{hatH}) 
and its supercharges (\ref{Qasuper}) by a unitary transformation generated 
by $\sigma_1$\,:
\be
\breve{\mathcal{H}}=\sigma_1\widehat{\mathcal{H}}\sigma_1\,,\qquad
\breve{\mathcal{S}}_1=\sigma_1\widehat{\mathcal{Q}}_1\sigma_1\,,\qquad
-\breve{\mathcal{S}}_2=\sigma_1\widehat{\mathcal{Q}}_2\sigma_1\,.
\ee
The commutator of $\widehat{\mathcal{H}}$ with $\sigma_1$  is
$[\widehat{\mathcal{H}},\sigma_1]=2\sigma_3\sigma_1$.
Denoting by $\tau$ the parameter associated
with evolution generated by $\widehat{\mathcal{H}}$,
we find the dynamical integrals corresponding to $\sigma_1$ and 
$\sigma_2=i\sigma_3\sigma_1$\,:
$\widehat{\Sigma}_a(\tau)=\exp(-i\widehat{\mathcal{H}}\tau)
\widehat{\Sigma}_a(0)
\exp(i\widehat{\mathcal{H}}\tau)=e^{-i2\sigma_3\tau}\widehat{\Sigma}_a(0)$,
$a=1,2$,
where  $\widehat{\Sigma}_a(0)=\Sigma_a$, and $\Sigma_a$ are defined
in  (\ref{Sigmadef}).
{}From here we also find that 
$\widehat{\mathcal{S}}_a(\tau)=e^{-4i\sigma_3\tau}\breve{\mathcal{S}}_a$ 
are dynamical integrals for $\widehat{\mathcal{H}}$.

Proceeding from the  supercharges 
$\widehat{\mathcal{Q}}_a$ and dynamical odd 
integrals $\widehat{\Sigma}_a$ and $\widehat{\mathcal{S}}_a$, 
we find  the symmetry of the system $\widehat{\mathcal{H}}$.
Besides the Hamiltonian  $\widehat{\mathcal{H}}$ and the listed 
fermionic  integrals, 
its set of generators  also includes  
bosonic time-independent integrals $\mathcal{Z}=-\frac{1}{4}\sigma_3$ and 
$\mathcal{I}=\frac{1}{4}\mathbb{I}$, and 
the even dynamical integrals of motion $\widehat{\mathcal{J}}_\pm(\tau)$
and $\widehat{\mathcal{C}}_\pm(\tau)$,
which have the same form as ${\mathcal{J}}_\pm$ 
and ${\mathcal{C}}_\pm$
 in (\ref{J+-t}) and (\ref{C+-t}) but 
with evolution parameter $t$ changed here for $\tau$.
The Hamiltonian of the extended system (\ref{Hextbr})
is related to supersymmetric Hamiltonian 
operators (\ref{hatH}) and (\ref{breveH})
by an equality
\be\label{HsusyHext}
\mathcal{H}=\frac{1}{2}(\widehat{\mathcal{H}}+
\breve{\mathcal{H}})=\widehat{\mathcal{H}}-\sigma_3\,.
\ee
In correspondence with this relation,
we introduce a notation 
for a linear combination of Hamiltonian 
$\widehat{\mathcal{H}}$ and bosonic integral
$\mathcal{Z}$,
\be\label{J0HZ}
\mathcal{J}_0=\frac{1}{4}\widehat{\mathcal{H}}+\mathcal{Z}\,.
\ee
As a result we find that 
the system given by the Hamiltonian operator (\ref{hatH})
is described by the same super-extended 
Schr\"odinger symmetry (\ref{supmat1})--(\ref{supmat6}) as the system
(\ref{Hextbr}) with the already identified
relation between the even generators,
while the correspondence between odd
generators of both systems   is given by
$\Sigma_a=\widehat{\Sigma}_a(0)$, 
and
\begin{equation}\label{Q+-Qa}
\mathcal{Q}_\pm(0)=\frac{1}{8}[\widehat{\mathcal{Q}}_1+
\widehat{\mathcal{S}}_1 \pm i(\widehat{\mathcal{Q}}_2-
\widehat{\mathcal{S}}_2)]|_{\tau=0},\qquad
{\mathcal{S}}_\pm(0)=\frac{1}{8}[\widehat{\mathcal{Q}}_2+
\widehat{\mathcal{S}}_2 \mp i(\widehat{\mathcal{Q}}_1-
\widehat{\mathcal{S}}_1)]|_{\tau=0}.
\end{equation}
Note that
both systems (\ref{hatH}) and  (\ref{Hextbr})
are characterized by the same number of
time-independent and dynamical integrals. 
The key difference is that coherently with different structure of
Hamiltonian operators related  by Eq. (\ref{HsusyHext}),
the system (\ref{Hextbr}) has two time-independent
integrals $\Sigma_a$ anticommuting for the central charge $\mathcal{I}$,
while time-independent integrals $\widehat{\mathcal{Q}}_a$ of
the system (\ref{hatH}) anti-commute for the Hamiltonian
operator $\widehat{\mathcal{H}}$.
Another essential difference is that the even dynamical integrals 
$\mathcal{J}_+$ and $\mathcal{J}_-$ of the system (\ref{Hextbr})
commute for the third generator $\mathcal{J}_0$
of $\mathfrak{sl}(2,\R)$ sub-algebra being the 
rescaled Hamiltonian $\mathcal{H}$, 
while the corresponding 
dynamical integrals $\widehat{\mathcal{J}}_+$ and 
$\widehat{\mathcal{J}}_-$ of the system (\ref{HsusyHext})
anticommute for a linear combination of its Hamiltonian
$\widehat{\mathcal{H}}$ and bosonic integral $\mathcal{Z}$.

We can apply to the symmetry generators 
of the system
described by the Hamiltonian $\widehat{\mathcal{H}}$ 
the same unitary transformation generated 
by nonlocal operator (\ref{Utrans})
and then realize the reduction of the transformed 
operators by means of projection
$\widetilde{\mathcal{X}_i}\mapsto  \Pi_+\widetilde{\mathcal{X}_i}\Pi_+$
to the proper eigenspace of eigenvalue $+1$ 
of $\sigma_3=\widetilde{\sigma_3\mathcal{R}}$.
In this way we, again, reproduce the  hidden superconformal 
symmetry of the single quantum harmonic oscillator
system. The peculiarity in this case is, however,
that  according to relation (\ref{HsusyHext}),
the application of unitary transformation and projection
to the Hamiltonian $\widehat{\mathcal{H}}$ results not in 
Hamiltonian of the harmonic oscillator operator $L$ 
but in a nonlocal operator\,: 
$\widehat{\mathcal{H}}\mapsto L-\mathcal{R}$.
\vskip0.1cm

In conclusion of this section it is worth noting that 
like the Darboux generators
(\ref{A0Darboux}) obtained from the 
ground state $\psi_0$,
the  operators (\ref{A-+0def}) constructed 
on the base of non-physical eigenstate $\psi_{-0}=1/\psi_0$
intertwine  $H_-$ and $H_+$ but with an additional shift\,:
\be\label{AAHH}
A^+_{-0}H_-=(H_+-4)A^+_{-0}\,, 
\qquad 
A^-_{-0}(H_+-4)=H_-A^-_{-0}\,.
\ee
It is because of such an additional shift matrix operators
$\breve{\mathcal{S}}_a$, unlike $\widehat{\mathcal{Q}}_a$,
correspond to dynamical integrals of motion of the system 
 $\widehat{\mathcal{H}}$.
 In the case of reflectionless and 
 finite-gap systems, there exist two pairs of
 operators that intertwine the corresponding partner systems
 exactly in   the same way, without additional shift.
 As a consequence, instead of two time-independent  supercharges,
 the  Darboux-extended systems are characterized there by 
 four time-independent supercharges,
which generate the corresponding non-trivial Lax-Novikov 
 integrals. For details see  ref. \cite{AraPly}.

\section{Two-step isospectral  Darboux chain}\label{Sec5}

The extended system  (\ref{Hextbr}) cannot be produced 
by a usual quantum mechanical supersymmetric construction of the form
(\ref{hatH}) based on some superpotential $W(x)$ that is equivalent 
to application of a one-step Darboux transformation. 
Let us show that it can be generated 
via a two-step isospectral  Darboux chain,
that uses a certain Jordan state of the quantum harmonic oscillator, 
with subsequent application of a 
simple limit procedure. 
The corresponding extended system obtained 
via the two-step isospectral Darboux chain
possesses a set of time-independent and 
dynamical integrals of motion.  After application of the limit
procedure these integrals give us  the generators of the 
super-extended Schr\"odinger symmetry 
of the system (\ref{Hextbr}). 

Consider a Darboux-Crum transformation
based on the seed states $\psi_0(x)$ 
and $\chi_0(x)$, where $\psi_0(x)$  is a normalized
ground state of the quantum harmonic oscillator (\ref{Lqhosc}) and 
$\chi_0(x;\mu)$ is its Jordan state of order two corresponding 
to the same energy $E=1$ \cite{CarPly},
\begin{equation}
\chi_0(x;\mu)=\mu\widetilde{\psi_0(x)}+\psi_0(x)\int_{0}^x\frac{1}{(\psi_0(t))^2}I_0(t)dt\,.
\end{equation}
Here  $\mu$ is a real constant, 
\begin{equation}
\label{tildepsi0}
\widetilde{\psi_0(x)}=\psi_0(x)\int_{0}^x(\psi_0(t))^{-2}dt\,
\end{equation}
is a  linear independent from $\psi_0(x)$ non-physical
eigenstate of $L$ of the same energy $E=1$,
and 
\be
I_0(x)=\int_{-\infty}^x (\psi_0(t))^2dt
\ee
is a monotonic function that varies between $0$ and $1$ 
as soon as the ground state wave function $\psi_0(x)$ is normalized for one. 
Note that the potential of the system shifted for corresponding eigenvalue $E=1$ 
is `extracted' from  $I_0(x)$ by  the Schwarzian\,: 
 $-\frac{1}{2}S(I_0)= x^2-1$,
where $S(f)=(f'''/f')-\frac{3}{2}(f''/f')^2$.
The application of the operator 
 $a^-$  to the state $\chi_0(x;\mu)$ produces the function
\begin{equation}\label{varphi0}
a^{-}\chi_0(x;\mu)=\frac{\mu+I_0(x)}{\psi_0(x)}=\varphi_{-0}(x;\mu)\,,
\end{equation}
that satisfies a relation $a^+\varphi_{-0}(x;\mu)=-\psi_0(x)$.
The state $\varphi_{-0}(x;\mu)$ 
is a linear combination of the non-physical eigenstate $\psi_{-0}(x)=1/{\psi_0(x)}$
of $L$ of eigenvalue  $-1$, that we used before 
to generate the system (\ref{breveH}),
and of a linear independent non-physical eigenstate 
$\widetilde{\psi_{-0}(x)}$  of $L$ of the same eigenvalue $-1$
constructed 
according to relation analogous to  (\ref{tildepsi0})\,:
\be
\varphi_{-0}(x;\mu)=\mu\psi_{-0}(x)+\widetilde{\psi_{-0}(x)}\,.
\ee
We choose the value of parameter $\mu$ in 
one of the infinite intervals
$(-\infty,-1)$ or $(0,\infty)$ for which  $\varphi_{-0}(x;\mu)$
is a nodeless on a real line function being 
a non-physical 
eigenstate of $H_+=a^-a^+$ of zero eigenvalue,
 $H_+\varphi_{-0}(x;\mu)=0$.
For   the wave function  $\chi_0(x;\mu)$ we find equivalently that it
satisfies a relation
$a^-a^+a^-\chi_0(x;\mu)=0$,  and therefore 
$(L-1)^2\chi_0=0$. This  means that $\chi_0(x;\mu)$  is indeed 
the Jordan state of order two of $L=H_-+1$ corresponding to eigenvalue $E=1$.

We can use the eigenstate $\varphi_{-0}(x;\mu) $ of $H_+$ 
as a seed state for a new Darboux transformation 
which produces the first order differential operators
\be
A^-_\mu=\varphi_{-0}(x;\mu)\frac{d}{dx}\frac{1}{\varphi_{-0}(x;\mu)}=
 \frac{d}{dx}+W(x;\mu)\, , \qquad 
A^+_\mu= (A^-_\mu)^\dagger\,,
\ee
where
\be
W(x;\mu)=-(\ln \varphi_{-0}(x;\mu))'=- x -\frac{\psi_0(x)}{\varphi_{-0}(x;\mu)}\,.
\ee
These operators factorize the Hamiltonian operators
$H_+=H_-+2$ and 
\be\label{H-muWron}
H_\mu=H_++2W'=H_--2\left(\ln (I_0(x)+ \mu)\right)''\,,
\ee
$A^+_\mu A^-_\mu=H_+$, $A^-_\mu A^+_\mu=H_\mu$,
and intertwine them, 
$A^-_\mu H_+=H_\mu A^-_\mu$, 
$A^+_\mu H_\mu=H_+ A^+_\mu$.
In (\ref{H-muWron}),  the argument of logarithm
is Wronskian of the states $\psi_0(x)$ and 
$\chi_0(x;\mu)$\,:  $Wr(\psi_0(x),\chi_0(x;\mu))=I_0(x)+ \mu$.
Considering the second order differential operators
given by a composition of the first order Darboux generators,
 \be\label{secondA}
 \A^-_\mu=A^-_\mu a^-\,,\qquad
 \A^+_\mu=a^+A^+_\mu\,,
 \ee
we find that they intertwine the Hamiltonian operators
$H_-=L-1$ and $H_\mu$, 
\be\label{A2inter}
\A^-_\mu H_-=H_\mu \A^-_\mu\,,\qquad
\A^+_\mu H_\mu=H_- \A^+_\mu\,,
\ee
and also satisfy relations $\A^+_\mu\A^-_\mu=(H_-)^2$,
$\A^-_\mu\A^+_\mu=(H_\mu)^2$.
By construction, $\ker(\A^-_\mu)=\text{span}\,\{\psi_0(x),\chi_0(x;\mu)\}$.
The Darboux-deformed oscillator system described by the 
Hamiltonian operator $H_\mu$ is \emph{completely isospectral}
to the system $H_-$. Its eigenstates 
with eigenvalues $E=2n$, $n=1,2\ldots$, are obtained
by mapping the  eigenstates $\psi_n(x)$ of the harmonic
oscillator by  the operator $\A^-_\mu$\,:
$\psi_n(x)\mapsto \psi_n(x;\mu)=\A^-_\mu\psi_n(x)$,
$H_\mu\psi_n(x;\mu)=2n\psi_n(x;\mu)$.
The (not normalized) ground state of zero energy of the system
$H_\mu$ is described by wave function 
$\psi_0(x;\mu)=\frac{1}{\varphi_{-0}(x;\mu)}$,
where $\varphi_{-0}(x;\mu)$ is a wave function (\ref{varphi0}).
It is obtained by application of the operator  $\A^-_\mu$ to 
the non-physical eigestate (\ref{tildepsi0}) of $H_-=L-1$ of zero energy,
  $\A^-_\mu\widetilde{\psi_0(x)}=-\psi_0(x;\mu)$.

Thus, we obtained the completely 
isospectral pair of the Hamiltonian operators $H_-$ and $H_\mu$, 
from which 
we compose the extended system described by the matrix 
Hamiltonian operator
\begin{equation}\label{calHmu}
\mathcal{H}_\mu=\left(
\begin{array}{cc}
H_\mu & 0   \\
 0 & H_-    
\end{array}\right).
\end{equation}
Its completely isospectral subsystems $H_-$ and 
$H_\mu$, as we have seen, 
are intertwined by the second order 
operators (\ref{secondA})
according to (\ref{A2inter}). 
On the other hand,
the first order operators $A_\mu^-$ and $A^+_\mu$ intertwine 
$H_+=H_-+2$ and $H_\mu$.
Therefore, these first order operators intertwine the subsystems
$H_-$ and $H_\mu$ of the extended system (\ref{calHmu})
but with a relative shift in comparison with 
(\ref{A2inter}), 
\be\label{Amuinter}
A^-_\mu H_-=(H_\mu-2) A^-_\mu\,,\qquad 
A^+_\mu (H_\mu-2)=H_- A^+_\mu\,.
\ee
 {}From this construction we have two Darboux schemes\,: 
the scheme based on the ground eigenstate and Jordan state
of the harmonic oscillator which produces the second order 
intertwining operators  $\A^\pm_\mu$. We denote such a scheme
 $(\psi_0(x),\chi_0(x;\mu))$. Another scheme is based on the 
 non-physical eigenstate $\varphi_{-0}(x;\mu)$ of $L$ of eigenvalue $-1$, 
 which we denote as $(\varphi_{-0}(x;\mu))$.
 We can construct one more Darboux scheme $(\psi_0(x),\psi_1(x), a^+\chi_0(x;\mu))$
 based on the physical eigenstates $\psi_0(x)$ and $\psi_1(x)$
 and the state $a^+\chi_0(x;\mu)$. This scheme gives rise 
  to the third order intertwining operators 
  $\mathcal{A}^-_\mu=A^-_\mu(a^-)^2=\A^-_\mu a^-$
  and $\mathcal{A}^+_\mu=(\mathcal{A}^-_\mu)^\dagger$.
  These operators also intertwine $H_-$ and $H_\mu$
  but with opposite  relative shift in comparison with (\ref{Amuinter})\,: 
  $\mathcal{A}^-_\mu H_-=(H_\mu+2)\mathcal{A}^-_\mu$,
    $\mathcal{A}^+_\mu (H_\mu+2)=H_-\mathcal{A}^+_\mu$.
 Using the intertwining operators of these three Darboux schemes,
 we construct three pairs of anti-diagonal (odd with respect to $\Gamma=\sigma_3$) operators
\begin{equation}
\mathcal{Q}_{\mu 1}=\left(
\begin{array}{cc}
0 & \A^- _\mu  \\
 \A^+_\mu & 0  
\end{array}\right),\quad 
\mathcal{Q}_{\mu 2}=i\sigma_3\mathcal{Q}_{\mu 1}\,,
\quad
\mathcal{S}_{\mu 1}=\left(
\begin{array}{cc}
0 & A^-_\mu   \\
 A^+_\mu & 0  
\end{array}\right),\quad 
\mathcal{S}_{\mu 2}=i\sigma_3\mathcal{S}_{\mu 1}\,,
\end{equation}
\begin{equation}
\mathcal{L}_{\mu 1}=\left(
\begin{array}{cc}
0 & \mathcal{A}^-_\mu   \\
 \mathcal{A}^+_\mu& 0  
\end{array}\right),\qquad
 \mathcal{L}_{\mu 2}=i\sigma_3\mathcal{L}_{\mu 1}\,.
\end{equation}
Using the relations 
between the intertwining operators 
$\A^-_\mu A^+_\mu=A^-_\mu a^-A^+_\mu$,
$A^+_\mu \A^-_\mu=(H_-+2) a^-$,
$\mathcal{A}^-_\mu A^+_\mu=A^-_\mu (a^-) A^+_\mu$,
$A^+_\mu \mathcal{A}^-_\mu=(H_-+2)(a^-)^2$,
we also construct diagonal (even) operators
\begin{equation}
\mathcal{C}_{\mu-}=\left(
\begin{array}{cc}
A^-a^-A^+ & 0   \\
 0 & (H_-+2)a^-
\end{array}\right),\qquad
\mathcal{J}_{\mu-}=\left(
\begin{array}{cc}
A^-(a^-)^2A^+ & 0   \\
 0 & (H_-+2)(a^-)^2
\end{array}\right),
\end{equation}
and Hermitian conjugate operators $\mathcal{C}_{\mu+}$ and 
 $\mathcal{J}_{\mu+}$.
 With respect to the Hamiltonian $\mathcal{H}_\mu$,
  the only pair of time-independent 
  integrals are the supercharges $\mathcal{Q}_{\mu a}$, $a=1,2$. 
  Other operators  have to be dressed
  with the unitary evolution operator $U(t)=\exp\left(i\mathcal{H}_\mu t\right)$\,:
  $\mathcal{S}_{\mu a}\mapsto U^{-1}(t)\mathcal{S}_{\mu a}U(t)$, etc.,
  that gives us the corresponding dynamical integrals of motion.
  The obtained in such a way time-independent and dynamical 
  integrals together with Hamiltonian operator $\mathcal{H}_\mu$ generate
  a kind of nonlinear superalgebra corresponding 
  to a nonlinear deformation of the super-Schr\"odinger 
  symmetry. 
  
  We are not interested here in explicit form
  of such a nonlinear superalgebra, but just note that
  when $\mu\rightarrow \pm \infty$, we have
   $(\ln(I(x) + \mu))' \rightarrow 0$.
   As a result, in any of the  two  limits
   the Hamiltonian $H_\mu$ transforms into $H_-$,
   and the matrix Hamiltonian transforms into extended Hamiltonian
   (\ref{Hextbr}) shifted for the minus unit matrix\,: 
   $\mathcal{H}_\mu\rightarrow \mathcal{H} -\mathbb{I}$.
   In this limit we also have 
$A^\pm_\mu\rightarrow -a^\mp$, and 
find that the constructed operators 
transform as follows\,:
\begin{eqnarray}
\mathcal{Q}_{\mu 1}\rightarrow -(\mathcal{H}-1)\sigma_1\,,&&\quad
\mathcal{Q}_{\mu 2}\rightarrow (\mathcal{H}-1)\sigma_2\,,\\
\mathcal{S}_{\mu a}\rightarrow-\breve{\mathcal{S}}_a\,,&&\quad
\mathcal{L}_{\mu a}\rightarrow-(\mathcal{H}-2+\sigma_3)
\widehat{\mathcal{Q}}_a\,,\\
\mathcal{C}_{\mu -}\rightarrow (\mathcal{H}-\sigma_3)\mathcal{C}_-\,,&&
\quad 
\mathcal{C}_{\mu +}\rightarrow \mathcal{C}_+(\mathcal{H}-\sigma_3)\,,\\
\mathcal{J}_{\mu -}\rightarrow (\mathcal{H}-\sigma_3)\mathcal{J}_-\,,&&
\quad 
\mathcal{J}_{\mu +}\rightarrow \mathcal{J}_+(\mathcal{H}-\sigma_3)\,.
\end{eqnarray}
In such a way we reproduce all the corresponding integrals
of the system  (\ref{Hextbr}) that generate the super-extended 
Schr\"odinger symmetry 
lying behind the hidden superconformal symmetries
$\mathfrak{osp}(1|2)$ and $\mathfrak{osp}(2|2)$
of a single quantum harmonic oscillator.

\begin{figure}[htbp]
\begin{center}
\includegraphics[scale=0.8]{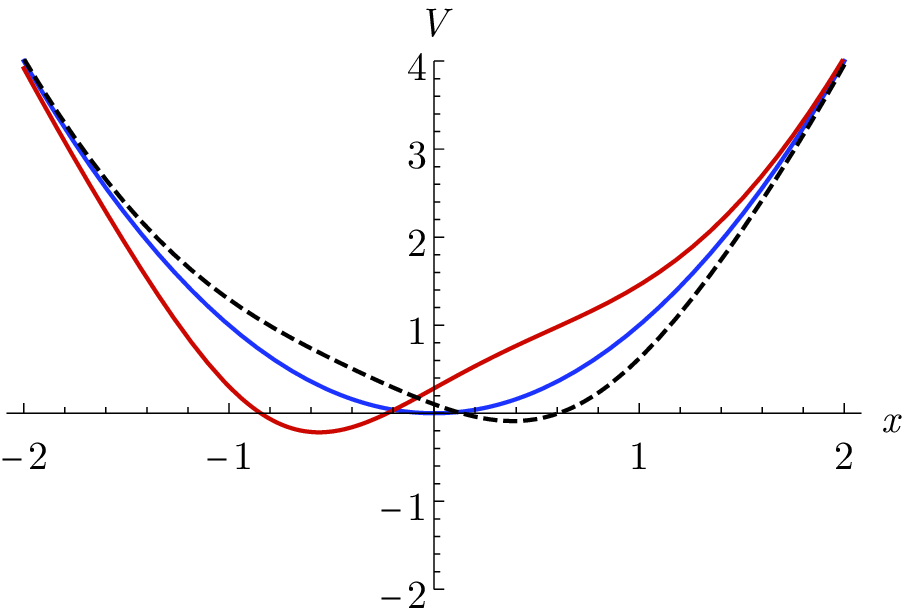}\includegraphics[scale=0.9]{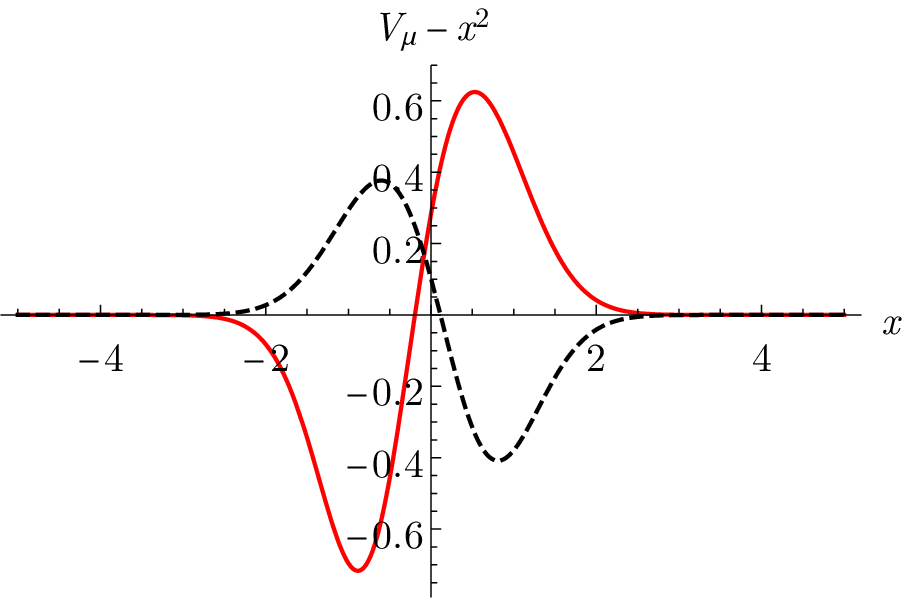}
\caption{ On the left: Isospectrally deformed potential $V_\mu$ 
at $\mu=1$  and $\mu=-3$ is shown by continuous red and dashed black
lines, respectively. 
On the right: The difference $V_\mu(x)-x^2$  given by the last term in Eq. (\ref{H-muWron}) 
is shown for the same values $\mu=1$  and $\mu=-3$.
With increasing value of modulus 
of the deformation parameter $\mu$ 
the amplitudes of minimum and maximum of  the difference $V_\mu(x)-x^2$ decrease, 
and in both  limits $\mu \rightarrow\pm \infty$
the deformed potential $V_\mu(x)$ transforms into harmonic potential $V=x^2$
shown on the left by continuous blue line.
}\label{Fig1}
\end{center}
\end{figure}

\begin{figure}[htbp]
 \begin{center}
\includegraphics[scale=0.4]{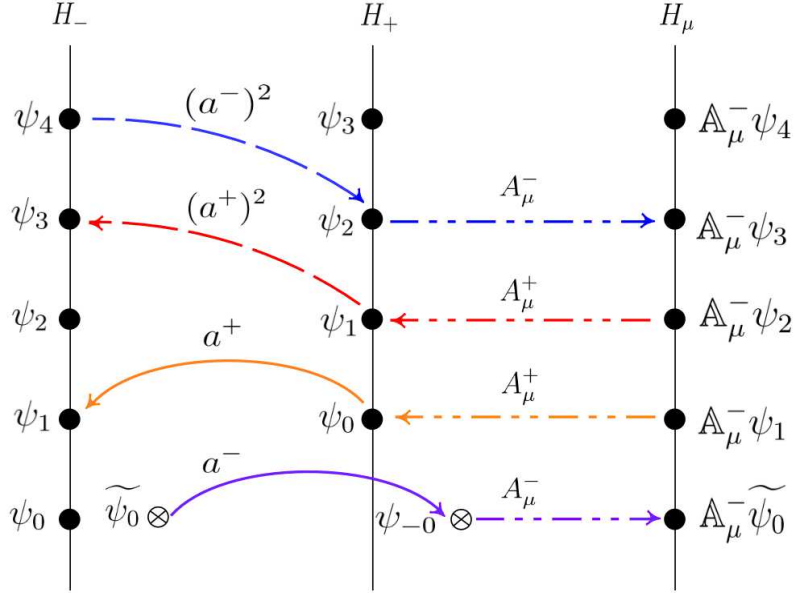}
\caption{Mapping of eigenstates of the systems $H_-$ and $H_\mu$  
by intertwining operators 
$\A^\pm_\mu$ and $\mathcal{A}^\pm_\mu$ 
via eigenstates of intermediate system $H_+$.
The ground state $\A^-_\mu\widetilde{\psi_0 }$
 of 
$H_\mu$ is obtained by applying  $\A^-_\mu$ to
non-physical eigenstate $\widetilde{\psi_0 }$
of $H_-$. It also can be generated  by a not shown here  action of  $\mathcal{A}^-_\mu$
on non-physical  eigenstate $\widetilde{\psi_1}$ of $H_-$ via
non-physical eigenstate $\psi_{-0}$ of $H_+$. }
\label{Fig2}
\end{center}
\end{figure}

The isospectral deformation $V_\mu(x)$  of the 
harmonic oscillator potential is illustrated by 
Figure \ref{Fig1}, while Figure \ref{Fig2} illustrates the action 
of the intertwining operators $\A^\pm_\mu$ and 
$\mathcal{A}^\pm_\mu$.

In conclusion of this section we note 
 that the Hamiltonian (\ref{calHmu}) 
and the second order  intertwining operators 
$\A^\pm_\mu$ can be presented 
in alternative  form which corresponds to the anomaly-free 
scheme of quantization of classical  systems with second-order
supersymmetry \cite{PlyuSchw}.
For this we introduce the quasi-amplitude \cite{Brezh}
\be
\Xi(x)=\sqrt{\psi_{-0}(x)\varphi_{-0}(x;\mu)}\,.
\ee
It is a square root of the product of two non-physical eigenstates  of eigenvalue 
$-1$ of the quantum harmonic oscillator $L$.
 The rescaled 
function $\Xi(x)/\sqrt{\mu}$ transforms in the limit $\mu\rightarrow \pm \infty$
 into the 
non-physical eigenstate $\psi_{-0}$.
This function satisfies Ermakov-Pinney  equation 
\be 
-\Xi''+(x^2+1)\Xi=\frac{1}{4\Xi^3}\,.
\ee
In terms of quasi-amplitude,  
the first order differential operators
\be\label{Axi}
A^-_\Xi=\Xi(x)\frac{d}{dx}\frac{1}{\Xi(x)}=\frac{d}{dx}-x-\mathcal{W}(x)\,,
\qquad
A^+_\Xi=(A_\Xi)^\dagger\,
\ee
can be defined,
where 
\be\label{calWsup}
\mathcal{W}(x)=\frac{1}{2\Xi^2(x)}=
\frac{1}{2}\left(\ln (I_0(x)+\mu)\right)'\,.
\ee
Then the Hamiltonian $H_\mu$  and the intertwining operator $\A^-_\mu$
can be presented in the form
\be
\mathcal{H}_\mu=A^-_\Xi A^+_\Xi +\mathcal{W}^2 -  2\mathcal{W}'\sigma_3\,,
\qquad 
\A^-_\mu=-(A^-_\Xi-\mathcal{W})(A^+_\Xi+\mathcal{W})\,.
\ee
Function $\mathcal{W}(x)$ in the anomaly-free scheme of quantization
plays a role of superpotential for corresponding  classical system with
second order supersymmetry \cite{PlyPara,KliPly,PlyuSchw}.

\section{Free particle limit}\label{Sec6}

It is interesting to apply 
a  limit of zero frequency to the constructions 
which have allowed us to clarify the 
nature of 
 the
hidden superconformal symmetry of the single 
quantum harmonic oscillator. 
In this way  we identify  a hidden superconformal symmetry of the 
free particle.  A nontrivial point in this procedure  is that 
it is necessary to make the appropriate 
different rescalings 
for different   
generators of super-Schr\"odinger symmetry 
of the extended quantum harmonic oscillator 
systems in order to obtain correctly their  
analogs for the corresponding (doubled) free particle
system.

To start, let us restore the frequency parameter $\omega$ 
in the Hamiltonian of harmonic oscillator $L$ and in its 
ladder  operators $a^\pm$ but maintaining the  Planck constant 
$\hbar=1$ and mass parameter $m=\frac{1}{2}$ as before.
We have 
\begin{equation}\label{Lomega}
L\rightarrow L_\omega=\omega\left(-\frac{1}{\omega}\frac{d^2}{dx^2}+\omega x^2\right)
=\frac{1}{2}\omega\{a_\omega^+,a_\omega^-\}\,,
\qquad
 a^\pm\rightarrow  a_\omega^{\pm}= \mp\frac{1}{\sqrt{\omega}}\frac{d}{dx}+\sqrt{\omega}x,
\end{equation}
and 
$[L_\omega,a_\omega^{\pm}]=\pm 2\omega a_\omega^{\pm}$,
$[L_\omega,(a_\omega^{\pm})^2]=\pm 2\omega( a^{\pm}_\omega)^2$.
We consider the application of the zero frequency limit
for the extended system (\ref{Hextbr}) and then make a comment 
for the case of the system (\ref{hatH}).
Applying  the zero frequency limit to the matrix Hamiltonian 
operator (\ref{Hextbr}) with restored according to (\ref{Lomega}) frequency, 
$\mathcal{H}_\omega$,
it is convenient to rescale it, and we obtain
\begin{equation}\label{H0H0}
\mathcal{H}_{0}=2\mathcal{H}_\omega|_{\omega \rightarrow 0}=
\left(
\begin{array}{cc}
  H_0 &  0 \\
 0 &   H_0     
\end{array}
\right),\qquad H_0=-\frac{1}{2}\frac{d^2}{dx^2}.
\end{equation}
The  nontrivial bosonic integrals for the 
system (\ref{H0H0}) are given by 
\begin{equation}
\mathcal{P}=i\sqrt{2\omega}(\mathcal{C}_+-\mathcal{C}_-)
|_{\omega \rightarrow 0}=\frac{1}{\sqrt{2}}\left(
\begin{array}{cc}
  p & 0    \\
 0 &   p     
\end{array}
\right),\qquad
p=-i\frac{d}{dx}\,,
\end{equation}
\begin{equation}
\mathcal{G}=\sqrt{\frac{2}{\omega}}(\mathcal{C}_++\mathcal{C}_-)
|_{\omega \rightarrow 0}=\frac{1}{\sqrt{2}}
\left(
\begin{array}{cc}
  x-2tp &  0 \\
 0 &   x-2tp     
\end{array}
\right),
\end{equation}
\begin{equation}
\mathcal{D}=\frac{i}{2}(\mathcal{J}_+-\mathcal{J}_-)
|_{\omega \rightarrow 0}=\frac{1}{4}
\left(
\begin{array}{cc}
  \{x,p\}-4tH_0 & 0  \\
 0 &   \{x,p\}-4tH_0    
\end{array}
\right),
\end{equation}
\begin{equation}
\label{K}
\mathcal{K}=\frac{1}{2\omega}[\mathcal{J}_++\mathcal{J}_- 
+\frac{2}{\omega}\mathcal{H}_\omega]|_{\omega \rightarrow 0}=\frac{1}{2}
\left(
\begin{array}{cc}
  x^2-2 t\{x,p\}+4 t^2H_0 & 0  \\
 0 & x^2-2 t\{x,p\}+4 t^2H_0   
\end{array}
\right).
\end{equation}
These are generators of spatial translations ($\mathcal{P}$),
of the Galilean boosts ($\mathcal{G}$), of dilatations 
($\mathcal{D}$), and special conformal transformations 
($\mathcal{K})$ of the doubled free particle system 
(\ref{H0H0}).
The linear combinations $\mathcal{Q}_+-\mathcal{Q}_-$ 
and $\mathcal{S}_+-\mathcal{S}_-$ of the dynamical
odd integrals of the system (\ref{Hextbr}) produce
 time-independent supercharges for the system 
(\ref{H0H0}), 
\begin{equation}\label{eta1}
\pi_1=i\sqrt{2\omega}(\mathcal{Q}_+-\mathcal{Q}_-)
|_{\omega \rightarrow 0}=\frac{1}{\sqrt{2}}\left(
\begin{array}{cc}
  0 & p    \\
 p &   0     
\end{array}
\right),
\end{equation}
\begin{equation}\label{eta2}
\pi_2=i\sigma_3\pi_1=i\sqrt{2\omega}(\mathcal{S}_+-\mathcal{S}_-)
|_{\omega \rightarrow 0}=\frac{1}{\sqrt{2}}\left(
\begin{array}{cc}
  0 & ip    \\
 -ip &   0     
\end{array}
\right),
\end{equation}
while the linear combinations $\mathcal{Q}_++\mathcal{Q}_-$ 
and $\mathcal{S}_++\mathcal{S}_-$ give us the 
dynamical odd integrals, 
\begin{equation}\label{zeta1}
\xi_1=\sqrt{\frac{2}{\omega}}(\mathcal{Q}_++\mathcal{Q}_-)|_{\omega \rightarrow 0}=\frac{1}{\sqrt{2}}
\left(
\begin{array}{cc}
  0 & x-2tp  \\
 x-2tp &   0     
\end{array}
\right),
\end{equation}
\begin{equation}\label{zeta2}
\xi_2=i\sigma_3\xi_1=\sqrt{\frac{2}{\omega}}(\mathcal{S}_+
+\mathcal{S}_-)|_{\omega \rightarrow 0}=\frac{i}{\sqrt{2}}
\left(
\begin{array}{cc}
  0 & x-2tp  \\
 -x+2tp &   0     
\end{array}
\right).
\end{equation}
Notice the difference in frequency factors before
the corresponding linear combinations in (\ref{eta1}), (\ref{eta2})
and (\ref{zeta1}), (\ref{zeta2}).
The time-independent even integrals $\mathcal{I}$ and $\mathcal{Z}$ and odd
integrals $\Sigma_a$, $a=1,2$,  of the system (\ref{Hextbr})
are also the integrals for the system (\ref{H0H0}).
All these operators generate 
the $\mathcal{N}=2$ super-Shr\"odinger algebra 
for the doubled free particle system  (\ref{H0H0})
with the following nontrivial (anti)-commutation  relations\,:
\begin{equation}\label{superS1}
[\mathcal{D},\mathcal{H}_0]=i\mathcal{H}_0\,,\qquad 
[\mathcal{D},\mathcal{K}]=-i\mathcal{K}\,,\qquad 
[\mathcal{K},\mathcal{H}_0]=2i\mathcal{D}\,,\qquad
[\mathcal{G},\mathcal{P}]=2i\mathcal{I}\,,
\end{equation}
\begin{equation}\label{superS2}
[\mathcal{H}_0,\mathcal{G}]=-i\mathcal{P}\,,\qquad 
[\mathcal{K},\mathcal{P}]=i\mathcal{G},\qquad 
[\mathcal{D},\mathcal{P}]=\frac{i}{2}\mathcal{P}\,,\qquad
[\mathcal{D},\mathcal{G}]=-\frac{i}{2}\mathcal{G}\,,
\end{equation}
\begin{equation}\label{superS3}
[\mathcal{D},\pi_a]=\frac{i}{2}\pi_a\,,\quad 
[\mathcal{D},\xi_a]=-\frac{i}{2}\xi_a\,,\quad 
[\mathcal{Z},\pi_a]=-\frac{i}{2}\epsilon_{ab}\pi_b\,,\quad 
[\mathcal{Z},\xi_a]=-\frac{i}{2}\epsilon_{ab}\xi_b\,,
\end{equation}
\be\label{superS3+}
[\mathcal{H}_0,\xi_a]=-i\pi_a\,,\qquad
[\mathcal{K},\pi_a]=i\xi_a\,,\qquad
\ee
\begin{equation}\label{superS4}
[\mathcal{Z},\Sigma_a]=\frac{i}{2}\epsilon_{ab}\Sigma_b\,,\qquad
[\mathcal{P},\pi_a]=-i\Sigma_a\,,\qquad 
[\mathcal{G},\xi_a]=i\Sigma_a\,,\qquad 
\end{equation}
\begin{equation}\label{superS5}
\{\Sigma_a,\pi_{b}\}=\delta_{ab}\mathcal{P}\,, \qquad 
\{\Sigma_a,\xi_{b}\}=\delta_{ab}\mathcal{G}\,,\qquad
\{\Sigma_a,\Sigma_b\}=2\delta_{ab}\mathcal{I}\,,
\end{equation}
\begin{equation}\label{superS6}
\{\pi_a,\pi_b\}=2\delta_{ab}\mathcal{H}_0\,,\qquad
\{\xi_a,\xi_b\}=2\delta_{ab}\mathcal{K}\,,\qquad
\{\pi_a,\xi_b\}=2\delta_{ab}\mathcal{D}+2\epsilon_{ab}\mathcal{Z}\,.
\end{equation}
Note that the superalgebra of the extended (doubled)
quantum harmonic system (\ref{Hextbr}) can be presented in the  
form (\ref{superS1})--(\ref{superS6}) with the following correspondence
between the generators\,: 
\begin{equation}
\mathcal{H}_0 \leftrightarrow \mathcal{J}_0-\frac{1}{2}(\mathcal{J}_++\mathcal{J}_-)\,,\qquad 
\mathcal{K} \leftrightarrow \mathcal{J}_0+\frac{1}{2}(\mathcal{J}_++\mathcal{J}_-)\,,
\qquad \mathcal{D}\leftrightarrow \frac{i}{2}(\mathcal{J}_+-\mathcal{J}_-) 
\end{equation}
\begin{equation}
\mathcal{P}\leftrightarrow \sqrt{2} i(\mathcal{C}_+-\mathcal{C}_-)\,,\qquad
\mathcal{G}\leftrightarrow \sqrt{2}(\mathcal{C}_++\mathcal{C}_-)\,,
\ee
\be
\pi_1 \leftrightarrow \sqrt{2}i(\mathcal{Q}_+-\mathcal{Q}_-)\,,\qquad \pi_2
\leftrightarrow \sqrt{2}i(\mathcal{S}_+-\mathcal{S}_-)\,,\qquad
\end{equation}
\begin{equation}
\qquad
\xi_1 \leftrightarrow \sqrt{2}(\mathcal{Q}_++\mathcal{Q}_-)\,,
\qquad \xi_2\leftrightarrow \sqrt{2}(\mathcal{S}_++\mathcal{S}_-)\,,
\end{equation}
and  with the same generators  $\mathcal{Z}$, $\mathcal{I}$  and $\Sigma_a$
for both systems.

In the case of the supersymmetric system (\ref{hatH}) after reconstruction 
of the frequency parameter we have 
\begin{equation}\label{Homegasuper}
\widehat{\mathcal{H}}_\omega=\frac{1}{4}\left(
\begin{array}{cc}
  L_\omega+\omega & 0    \\
 0&   L_\omega -\omega    
\end{array}
\right).
\end{equation} 
The zero frequency limit applied to this
system produces the same 
doubled free particle system (\ref{H0H0}).
This can also be understood by noting that
after reconstruction of frequency, the superpotential 
in (\ref{hatH}) takes a form $W_\omega=\omega x$.
In the zero frequency limit $W_\omega\rightarrow 0$ and 
$\widehat{\mathcal{H}}_\omega=-\frac{d^2}{dx^2}+W_\omega^2+
W'_\omega\sigma_3\rightarrow -\frac{d^2}{dx^2}\I$.

The application of the unitary nonlocal operator
(\ref{Utrans}) to the odd generators $\pi_a$ and $\xi_a$ 
transforms  them into operators of  the diagonal form, and,
particularly, $\pi_1$ and $\xi_1$ take the form of the operators
$\mathcal{P}$ and $\mathcal{G}$. 
The transformed $\mathcal{P}$ and $\mathcal{G}$  take the 
anti-diagonal form, $\widetilde{\mathcal{P}}=\pi_1$,
$\widetilde{\mathcal{G}}=\xi_1$.
After projection to the proper subspace of eigenvalue $+1$ of 
$\sigma_3$, 
$\widetilde{\mathcal{X}_i}\mapsto  \Pi_+\widetilde{\mathcal{X}_i}\Pi_+$,
we left 
with a single free particle system $H_0$,
whose hidden superconformal $\mathfrak{osp}(1|2)$ symmetry
 is generated by local even integrals
 $H_0$, $D= \frac{1}{4}(\{x,p\}-4tH_0)$,
 $K=\frac{1}{2}x^2-t\{x,p\}+2t^2H_0$,
 and by odd integrals $P=p$, $G=x-pt$,
 with reflection operator $\mathcal{R}$ playing the role
 of the $\Z_2$-grading operator.
 The inclusion of nonlocal even integral 
 $Z=-\frac{1}{4}\mathcal{R}$  and nonlocal odd integrals  
  $i\mathcal{R}P$, $i\mathcal{R}G$ extends the hidden
  superconformal $\mathfrak{osp}(1|2)$ symmetry for the hidden
  superconformal $\mathfrak{osp}(2|2)$ symmetry
  of the free particle. Explicit form the (anti)-commutation
  relations can  easily be identified from the 
  corresponding relations from 
  (\ref{superS1})--(\ref{superS6}).
 
 \section{Discussion and outlook}\label{Sec7}

 The extended doubled quantum harmonic oscillator system 
 (\ref{Hextbr})  is not described by a usual construction of supersymmetric quantum mechanics. 
 Nevertheless it is possible to obtain it in the following way by starting 
 from  the level of classical mechanics.
Consider a classical system described by 
a Hamiltonian  
\be\label{Comm1}
H=p^2+W^2 +W'[\theta^+,\theta^-]
\ee
 with superpotential 
$W(x)=\sqrt{x^2+c^2}$, where $c>0$ is a constant and 
$\theta^+$ and  $\theta^-=(\theta^+)^*$  are Grassmann variables being 
classical analogs of the fermion creation-annihilation 
operators with a nonzero Poisson bracket 
$\{\theta^+,\theta^-\}_{{}_{PB}}=-i$.
 After quantization the odd variables $\theta^\pm$  transform into 
 fermionic creation-annihilation operators
 which can be realized in terms of  $\sigma$-matrices\,:
 $\theta^\pm\rightarrow \sigma_\pm=\frac{1}{2}(\sigma_1\pm i\sigma_2)$.
 A direct quantum analog of this system is a supersymmetric
 completely isospecral pair in the phase of spontaneously broken supersymmetry,
 with nonsingular superpartner potentials $V_\pm=x^2+c^2\pm x/\sqrt{x^2+c^2}$.
 The spectrum of subsystems is different from that 
 of the quantum harmonic  oscillator.
On the other hand, if before the quantization we realize a canonical 
transformation 
$x\rightarrow X=x+ N \partial G(x,p)/\partial p$,
$p\rightarrow P=p - N\partial G(x,p)/\partial x$, 
$\theta^\pm\rightarrow \Theta^\pm=e^{\pm iG(x,p)}\theta^\pm$,
where $N=\theta^+\theta^-$ and 
$G=\frac{1}{2}\arcsin \big((p^2-x^2-c^2)/(p^2+x^2+c^2)\big)$,
we obtain the canonically equivalent form of the 
Hamiltonian $H=P^2+X^2 +c^2$. In the canonically transformed system, the new 
classical Grassmann variables $\Theta^\pm$ completely decouple and are the odd 
integrals
of motion with Poisson bracket $\{\Theta^+,\Theta^-\}_{{}_{PB}}=-i$. 
The quantization of the canonically transformed system 
gives us exactly the  extended quantum system (\ref{Hextbr}) shifted just for the additive 
constant $c^2$.

The  classical system 
given  by the Hamiltonian $H=p^2+W^2 +2W'[\theta^+,\theta^-]$
with a simply  changed in comparison with (\ref{Comm1})  boson-fermion coupling constant in the 
last term is characterized at the classical level
by the second-order supersymmetry \cite{PlyPara,KliPly}. 
The application of the anomaly-free scheme of quantization  \cite{PlyuSchw} 
to such a system with superpotential (\ref{calWsup}) produces
the extended quantum system (\ref{calHmu}).
 
In Section  \ref{Sec4}, we generated superconformal symmetry
via dual Darboux transformations.
The analogous construction can be applied for investigation 
of superconformal symmetry in supersymmetric  
rationally extended  quantum harmonic and isotonic oscillator systems \cite{CIP}.
The results of such investigation 
will be presented by us elsewhere.

We derived the hidden superconformal symmetry of the
one-dimensional   quantum harmonic oscillator and clarified 
its nature.
It would be interesting to analyze from the same perspective 
the case of $d$-dimensional quantum harmonic oscillator systems
which also are characterized by the corresponding 
generalizations of the  hidden superconformal symmetry 
considered here \cite{CroRit,BalSchBar,BCLW,BDH2,BCM}. 

It is interesting to note that our construction of 
the one-parametric completely isospectral
deformation of the quantum harmonic oscillator 
from Section  \ref{Sec5} can be generalized
by taking instead of  the pair ($\psi_0,\chi_0(\mu)$)
of the seed states in the Darboux-Crum transformation
the pair ($\psi_n,\chi_n(\mu))$,
where $\chi_n(\mu)=\chi_n(x;\mu)$ is the Jordan state 
corresponding to eigenvalue $E_n$ \cite{CarPly}.
In this case we shall obtain 
completely isospectral deformations of the 
harmonic oscillator potential of a more complicated 
form. The properties of the 
quantum systems  obtained in such a way deserve a further investigation


\vskip0.2cm

\noindent {\large {\bf Acknowledgements} } 
\vskip0.4cm

LI acknowledges the CONICYT scholarship 21170053.
MSP acknowledges support from research project
USA1555.


\end{document}